\documentclass[%
 aip,
 amsmath,amssymb,
 reprint,%
]{revtex4-1}

\usepackage{amsmath,amsfonts,amssymb}
\usepackage{physics}
\usepackage{graphicx}
\usepackage{setspace}
\usepackage{tocloft}
\usepackage{float}
\usepackage[dvipsnames]{xcolor}

\cftpagenumbersoff{figure}
\cftpagenumbersoff{table} 

\usepackage{graphicx}% Include figure files
\usepackage{dcolumn}% Align table columns on decimal point
\usepackage{bm}% bold math
\usepackage[utf8]{inputenc}
\usepackage[T1]{fontenc}
\usepackage{mathptmx}
\usepackage{etoolbox}

\makeatletter
\def\@email#1#2{%
 \endgroup
 \patchcmd{\titleblock@produce}
  {\frontmatter@RRAPformat}
  {\frontmatter@RRAPformat{\produce@RRAP{*#1\href{mailto:#2}{#2}}}\frontmatter@RRAPformat}
  {}{}
}%
\makeatother
\begin{document}

\preprint{AIP/123-QED}

\title{Towards digital phantoms: emulating scattering with a spatial light modulator}
% Force line breaks with \\

\author{Kelsey Everts}
 \affiliation{School of Physics, University of the Witwatersrand, Private Bag 3, Wits 2050, South Africa}

\author{Cade Peters}
 \affiliation{School of Physics, University of the Witwatersrand, Private Bag 3, Wits 2050, South Africa}

\author{Andrew Forbes}
%\email[email:]{andrew.forbes@wits.ac.za}
\affiliation{School of Physics, University of the Witwatersrand, Private Bag 3, Wits 2050, South Africa}
\email[Corresponding author: ]{andrew.forbes@wits.ac.za}

\date{\today}% It is always \today, today,
             %  but any date may be explicitly specified

\begin{abstract}
\noindent The distortion of light's degrees of freedom when passing through complex random media is of great interest across a diversity of fields, e.g., scattering in biological studies. Emulating such media in a controlled laboratory setting conventionally relies on real-world physical samples (e.g., white paint), inhomogeneous mixtures with embedded scatterers, or biological tissue-mimicking phantoms. Such methods, while effective in certain contexts, are not without complexity and limitations: the exact medium properties are challenging to control and often require laborious preparation, external characterisation techniques, are not easily reproducible between studies and cannot be matched precisely by numerical simulations. Here, we propose a simple all-digital implementation of random scattering which can be readily implemented on any setup capable of producing digital holograms. Our approach employs binary random phase masks encoded onto a spatial light modulator which perturbs the input beam's phase and amplitude. We highlight two methods to precisely tune distortion strengths which show excellent agreement between simulated and measured results. We demonstrate distortion strengths comparable to real-world scattering samples and illustrate two example applications to emulate scattering of scalar and vectorial structured light. Finally we showcase the versatility of this toolkit for emulating various amplitude and phase profiles and suggest several easy to implement alternative modalities accessible with this method. This digital phantom circumvents many of the practical challenges of physical samples, making it ideally suited for applications at the intersection of structured light, biological imaging and optical communications.
\end{abstract}

\maketitle

\section{Introduction}

\begin{figure*}[htbp]
    \centering
    \includegraphics[width=0.7\linewidth]{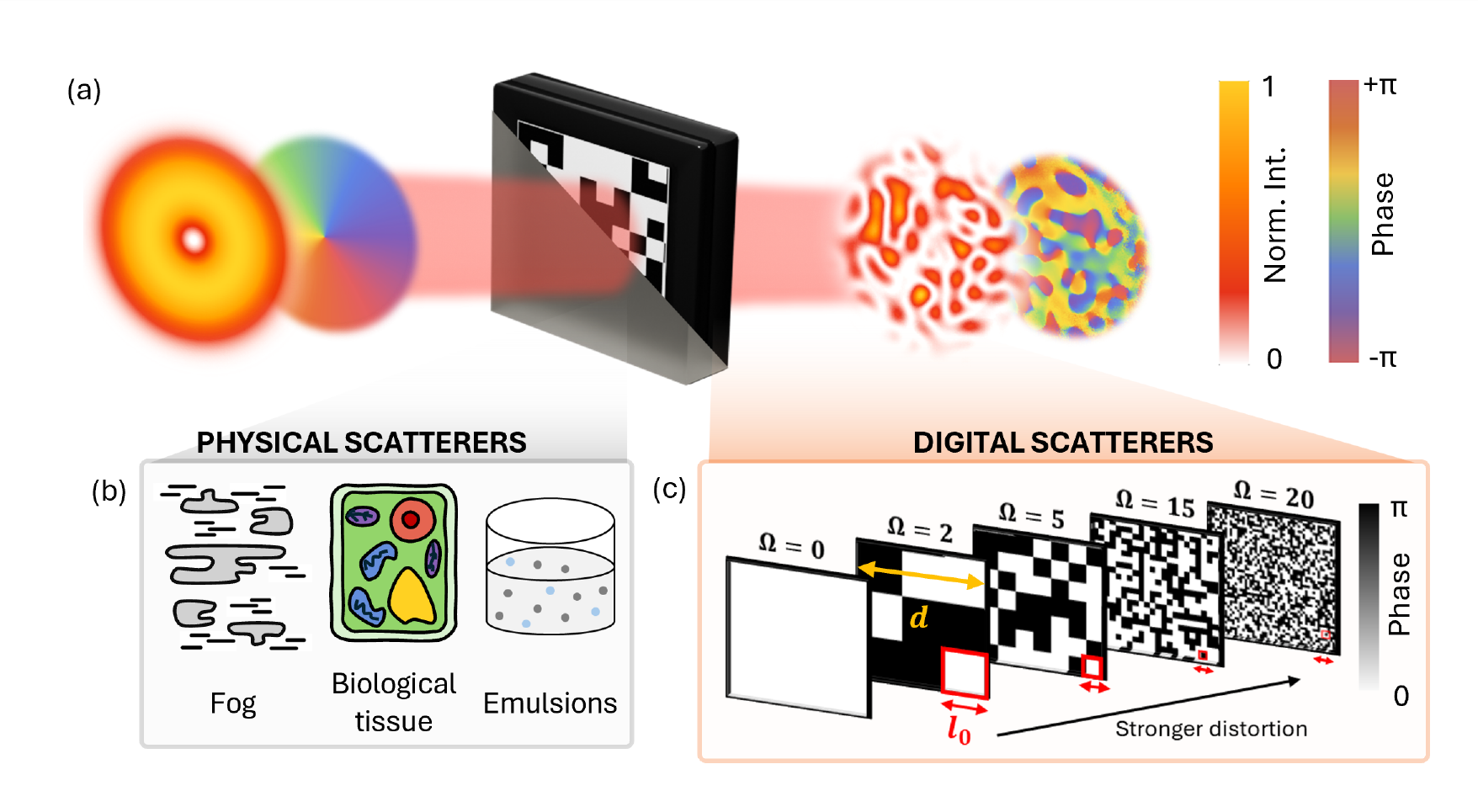}
    \caption{(a) Digitally simulated scattering mimicking the intensity and phase distortions imparted onto a pristine input beam by scattering samples. (b) Commonly used physical scattering samples. (c) Digital scattering masks: binary random phase masks encoded onto a spatial light modulator. Input parameters are the beam diameter, $d$, and the medium's transverse correlation length, $l_0$. Increasing distortion strengths are achieved by decreasing the encoded $l_0$ which determines the block size as highlighted in red.
    \label{fig:concept}}
\end{figure*}

\noindent Distortion is virtually unavoidable when propagating light through real-world channels. The scattering of light in complex channels such as biological tissue \cite{yoon2020deep,ntziachristos2010going}, underwater \cite{zeng2016survey,sahu2018theoretical}, optical fibre \cite{cao2023controlling} and through the atmosphere \cite{cox2020structured,peters2025structured} affects almost all of light's degrees of freedom leading to aberrated intensity profiles, scrambled phase profiles,  and speckle patterns\cite{Rotter2017mesoscopic,Gigan2022roadmap} as shown by the scrambled output phase and intensity profiles in Figure \ref{fig:concept}(a). Additionally, depolarisation \cite{He2021polarisation}, and pulse width lengthening \cite{cochenour2013temporal} can occur. Together, these decrease spatial resolution and contrast in imaging and sensing \cite{Bertolotti2022imaging} and increase inter-modal crosstalk \cite{katz2011focusing,sheng2007introduction}. This is detrimental in a multitude of applications including biological imaging and disease diagnosis \cite{yaqoob2008optical,ding2024wavefront}, where images are highly distorted or resolution is harshly decreased, optical communications through free and optical fibre where it increases bit error rates and decreases bandwidth \cite{essiambre2010capacity,richardson2013space}, and optical tweezing and micromanipulation \cite{jones2015optical}.  Understanding light's behaviour through these channels is therefore of tremendous interest \cite{cao2022shaping,lib2022quantum,Meglinski2024phase}, with the potential to improve the effectiveness of current imaging techniques and increase the rate at which we send and receive information.

Current methods for studying the effects of scattering media on light fields rely heavily on the use of physical samples. These range from ground glass \cite{bertolotti2012non}, parafilm \cite{courme2023manipulation}, ZnO nanoparticles \cite{boniface2017transmission,valzania2023online} and white paint \cite{vellekoop2007focusing,popoff2010measuring} to suspended scatterers in emulsion \cite{ye2021detection} and tissue mimicking phantoms \cite{chue2019optical} as shown in Figure \ref{fig:concept}(b). These methods, while forming effective scatterers, lack precise control over the distortion profile and severity, requiring external means to measure and calibrate these important properties \cite{hassaninia2017characterization} and making it challenging to reproduce results across studies. Traditional tissue-mimicking phantoms do allow for some degree of control over the medium's properties, but involve sophisticated equipment and time-consuming fabrications procedures \cite{krauter2015optical}, making them cumbersome to implement, even in well equipped environments. Additionally these require external means of characterization before use \cite{moffitt2006preparation}, degrade over time \cite{cabrelli2017stable} and are still limited in how precisely one can control their key properties \cite{pogue2006review}. 

The rise in popularity of structured light has led to the development of an advanced digital toolkit capable of tailoring almost all of lights degrees of freedom (DoFs) including phase \cite{yao2011orbital}, amplitude \cite{scholes2020structured}, polarization \cite{zhan2009cylindrical} and temporal profile \cite{chong2020generation}. These advances have not only made it possible to generate \cite{arrizon2007pixelated,ngcobo2013digital,shen2022generation} and detect \cite{forbes2016creation} many forms of structured beams in both the classical and quantum regimes, but also allow one to manipulate their profiles to replicate the effects of a variety of channels, all achievable with simple optical setups combined with complex amplitude shaping digital holograms \cite{Bromberg2014generating,Bender2018speckle}.

 Here, we propose the creation of digital phantoms by leveraging the versatility of the digital structured light toolkit to replicate the effects of scattering media. By making use of computer generated holograms, our approach avoids entirely the onerous fabrication procedures necessary for physical phantoms, allowing one to create a simulated channel in mere moments. We make use of binary phase masks as an example, alternating between 0 and $\pi$ to induce both phase and amplitude distortions onto the incident beam upon imaging or propagation. We test the effectiveness of these masks on both scalar and vectorial forms of structured beams, and compare their performance to analytically predicted behaviour and real-world scatterers, showing remarkable agreement in both. A key benefit of this approach is that the hologram can be used both numerically and experimentally, allowing one to closely simulate the experimental conditions with high fidelity, which we show for scalar and vectorial beams. Our approach allows for the rapid and reproducible replication of scattering media in a manner that is more accessible, easier to implement and directly comparable to simulation. We foresee our approach allowing for a vast increase in the ability to investigate scattering channels and allowing for more controlled and systematic studies of its effects on the many properties of optical fields, with a multitude of applications such as medical imaging and free space and optical fibre communications, both classical and quantum. 

\section{Concept}
%\cite{Rotter2017mesoscopic}. Consequently, wimposing random distortions and experimentally simulated thin, inhomogeneous phase and amplitude distorting media where other degrees of freedom are not distorted. 
In order to replicate the effects of the scattering media using a digital device such as a spatial light modulator (SLM), an appropriate choice of hologram must be determined. We illustrate an example in Figure \ref{fig:concept} (a), where we opt to use a hologram that is made up of super-pixel square blocks with spatially varying random phases of 0 or $\pi$, although any shape geometry and phase change is possible (see later sections). The effect of these elements is to delay or advance only specific areas of the beam's spatial profile, similar to real-world biological tissue, fog or suspended scatterers, shown in Figure \ref{fig:concept} (b). Once the beam has passed through this hologram, the now aberrated spatial profile will distort in both phase and amplitude as it propagates, just as the beam would if it had passed through a real-world scattering sample.  For simplicity we choose to focus on the thin medium limit allowing its approximation by a single phase screen, but stress that the approach we outline here can easily be extended to multiple scattering (thick) samples by multiple bounces off the SLM \cite{klug2023robust,peters2025structured}. 

We relate the strength of our digital distortion to that of real media through the transverse correlation length, $l_0$, the average distance over which the phase is correlated \cite{Bachmann2024universal}.  In our digital emulator this is the length of each phase block, as shown in Figure \ref{fig:concept} (c). While $l_0$ defines how disordered the medium is, the severity of the distortion experienced by an incident paraxial beam of light also depends on the transverse size of the beam. We therefore quantify the distortion strength of the medium as the unitless parameter 
\begin{equation}
    \Omega = d/l_0\,,
    \label{eq:disortion_strength}
\end{equation}
where $d$ is the input diameter of the incident beam defined according to the second moment beam radius. 

We illustrate the effect of varying $l_0$ visually in Figure \ref{fig:concept} (c). Intuitively, if $l_0>d$ ($\Omega<1$), the beam will experience a uniform phase and thus will not be distorted. However, if $l_0<d$ ($\Omega>1$) then there will be random phase flips across the beam profile, a digital distortion that is easily increased or decreased by using smaller or larger values values of $l_0$ for the same beam diameter $d$. Larger values of $\Omega$ correspond to smaller blocks of uniform phase and thus more rapidly varying phase fluctuations.

\section{Experimental Implementation}

To demonstrate our approach of simulating scattering media using digital holograms, we employed the experimental setup depicted in Figure \ref{fig:setup} (a). Here, a horizontally polarized Helium-Neon laser beam (wavelength $\lambda=633$~nm) was magnified to overfill and evenly illuminate an SLM screen using a $20\times$ objective lens $L_1$ and a collimating lens, $L_2$ ($f = 300$~mm). The mode(s) of interest were programmed onto the SLM using a complex amplitude modulation scheme \cite{arrizon2007pixelated}, with an example of the holograms for Laguerre Gaussian (LG) modes with azimuthal index $\ell=3$ and embedded Gaussian beam waist of $w_0 = 0.4$~mm shown in the inset of Figure \ref{fig:setup}(b) (bottom row) for several $\Omega$ along with the phase mask (top row). The flexibility of the hologram modulation scheme also allowed for programming flattop beams and Hermit Gaussian (HG) beams \cite{peters2025structured}.

\begin{figure*}[htbp]
    \centering
    \includegraphics[width=0.8\linewidth]{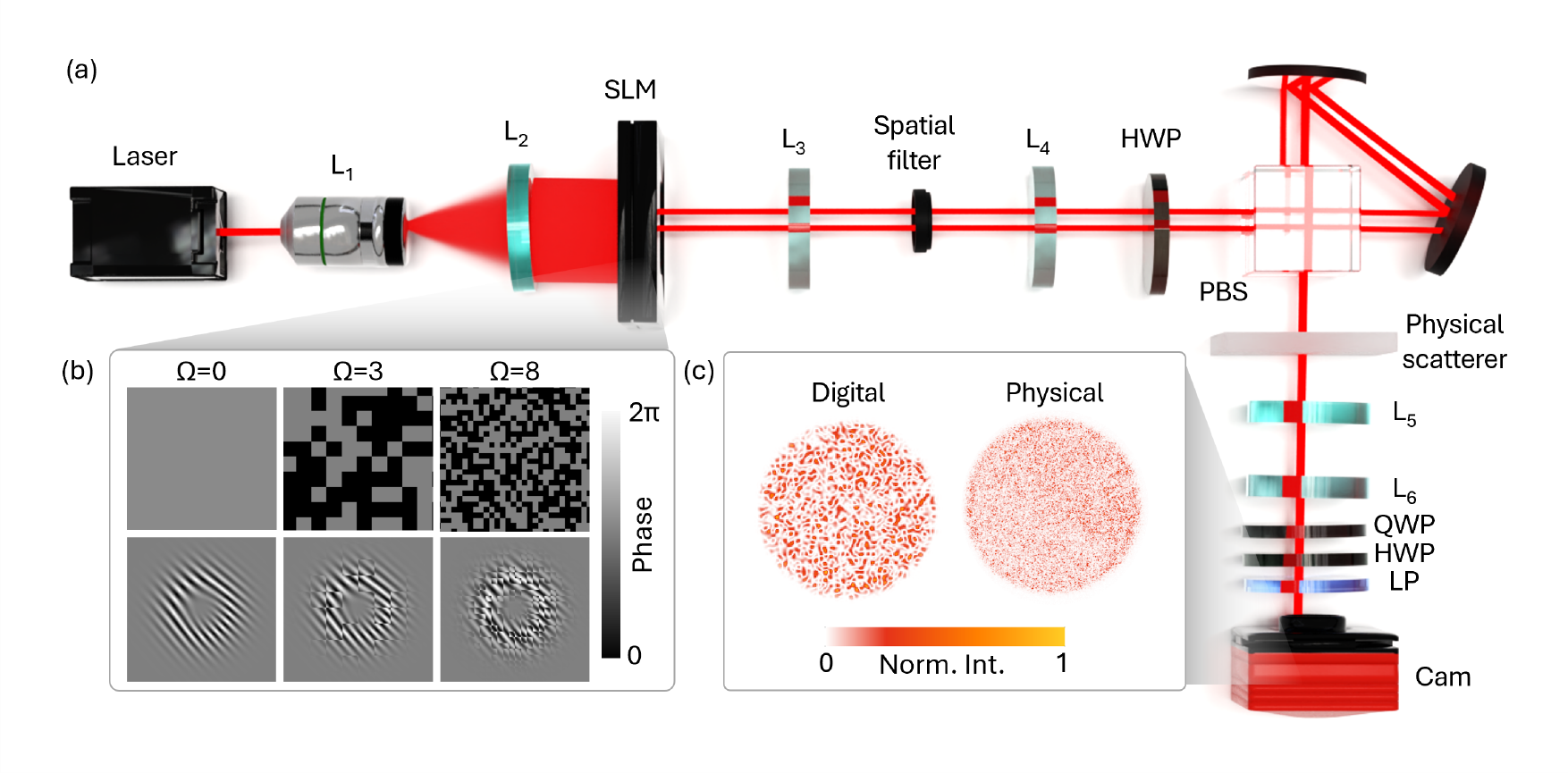} \label{fig:setup}
    \caption{(a) Experimental setup used to generate structured light modes and digitally simulated scattering consisting of an SLM imaged through a Sagnac interferometer onto a camera. (b) Top row: Phase masks for various distortion strengths, $\Omega$. Bottom row: Holograms encoded on the SLM for $\text{LG}_{\ell=3}^{p=0}$ mode superimposed with the corresponding digital phase mask. (c) Intensity images captured by the camera either of a digital or physical sample (parafilm) inducing aberrated flattop beam with similar profiles.
    } 
    \label{fig:setup}
\end{figure*}

For the purposes of performing full field measurements and generating vectorial structured light, we divided the SLM screen in half and encoded holograms independently on each half. The plane of the SLM was then imaged using a 4f imaging system $L_3$ and $L_4$ with$f_3=f_4=300$~mm. A variable size aperture was placed in the Fourier plane of $L_3$ to act as a spatial filter, isolating the first diffraction order of the SLM hologram where the desired complex field was encoded. A half-wave plate (HWP) was placed directly after $L_4$ converting them both to diagonal polarisation before the two beams (one from each half of the SLM) pass through a Sagnac interferometer \cite{shen2022generation}. The Sagnac interferometer consisted of a polarizing beam splitter (PBS) and two mirrors, allowing for the formation of a coaxial and collinear vectorial combination of the two beams. When taking data through physical scatterers, the samples were placed in the focal plane of $L_4$. The beams were then imaged from this plane onto a camera using another 4f imaging system consisting of lenses $L_5$ and $L_6$ ($f_5=f_6=300$~mm). A half-wave plate (HWP), quarter-wave plate (QWP) and linear polariser (LP) were placed in front of the camera to perform a full Stokes polarimetry measurement of the incident field. This allowed for a full characterization of any generated vectorial fields \cite{singh2020digital} as well as the phase retrieval of scalar beams \cite{dudley2014all}.

We show example intensity measurements of an aberrated flattop beam after passing though a digital and physical scatterer (parafilm) in Figure \ref{fig:setup} (c). Results for the digital scatterers were taken with the physical samples removed from the beam line and results of physical scatterers were taken with no binary phase mask encoded on the SLM holograms.

\section{Results}
\subsection{Tuning digital distortion strength}\label{sec:tuneable_distortion}

\begin{figure*}[htbp]
    \centering
    \includegraphics[width=0.7\linewidth]{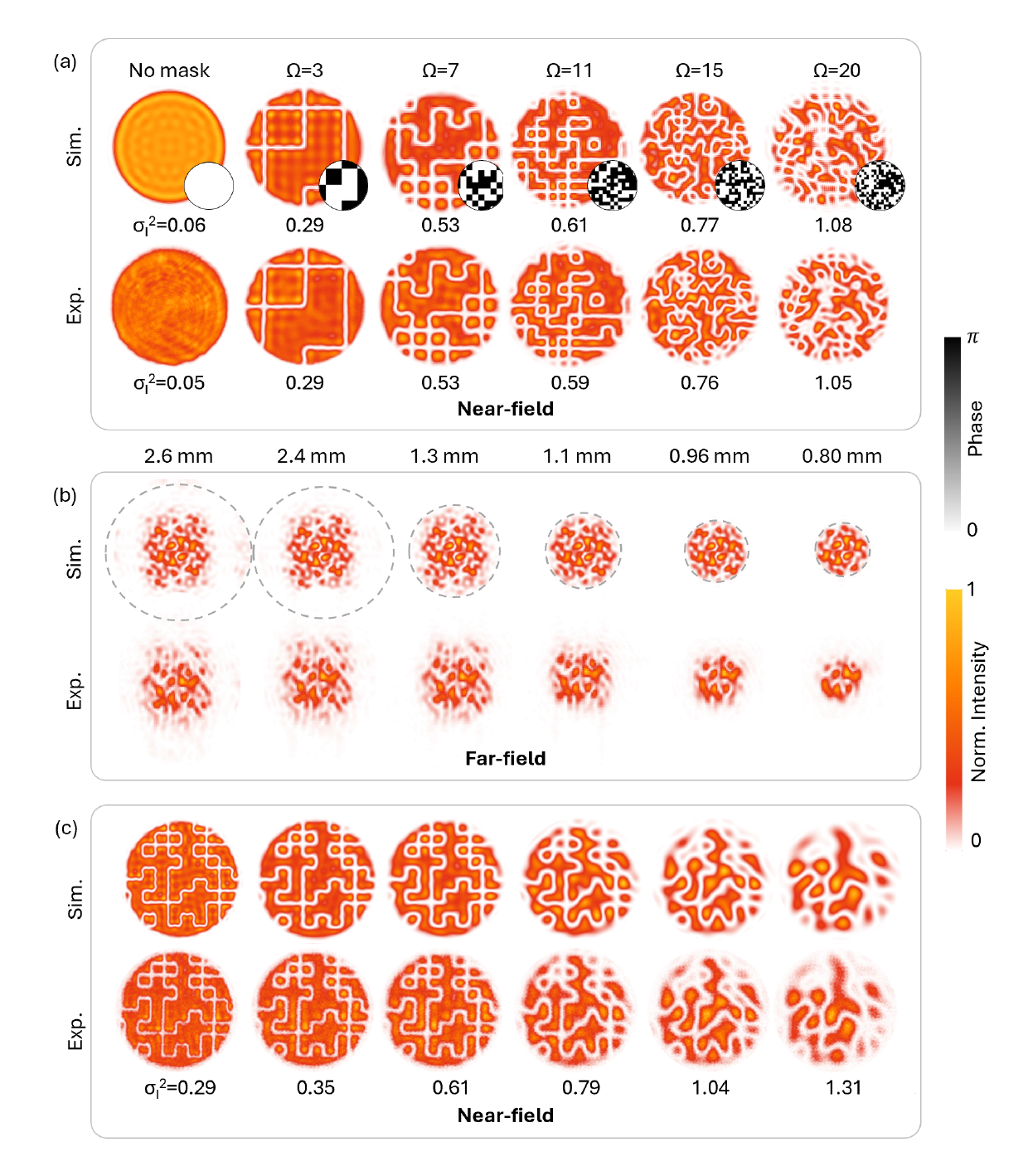}
    \caption{(a)  Near-field simulated and experimentally measured intensity profiles for different distortion strengths, $\Omega$ for one random phase mask realization (inset). Scintillation index, $\sigma_I^2$ shown below each beam. (b) Far-field diffraction patterns for fixed $\Omega =10$ with different spatial filter diameter in the far field. Dotted circles show aperture size. (c) Corresponding near-field images with varying intensity null thicknesses.
    }
    \label{fig:spatial_filter}
\end{figure*}

Due to the use of digital holograms in combination with a spatial filter and a finite apertured optical system, we are able to tune the nature of the distortion in two main ways. The first method is illustrated in Figure \ref{fig:spatial_filter} (a). This involves changing the encoded distortion strength parameter $\Omega$. We show the binary phase masks as insets as well as simulated and experimental intensity profiles of a flattop beam after impinging on the mask and passing through the first imaging system $L_3$ and $L_4$. We see that in the case when no mask is applied, we observe an round beam with approximately even intensity as expected. When we encode a mask with $\Omega > 0$, we observe sharp lines in the beam's intensity. These lines manifest due to the high spatial frequencies present in the beam due to the sharp phase jumps from 0 to $\pi$ at the boundaries between superpixels in the phase mask. As the beam propagates through the imaging system, these spatial frequencies are not captured by the lenses, resulting in null intensity regions in the beam at the image plane. Because these lines manifest at the interface between super pixels where the phase jumps between $0$ and $\pi$, we are able to directly correlate the null intensity lines with profile of the encoded mask in the near field, even though the encoded perturbation is phase-only at the SLM. We therefore observe an increase in the number of intensity null lines as $\Omega$ increases, resulting in a more severe perturbation to the field's amplitude structure in the  image plane. We quantified the severity of these intensity fluctuations using the scintillation index $\sigma_I^2$ which clearly increases as $\Omega$ increases \cite{Andrews1999scintillation}. The experimental and simulated intensity profiles and computed $\sigma_I^2$ values match very closely as shown in Figure \ref{fig:spatial_filter} (a) when using the same mask. This directly demonstrates the power of our approach over physical scattering samples where achieving such controlled random scattering, and such excellent agreement between simulation and experiment is not feasible.

The first method demonstrated how tuning $\Omega$ tunes the number of intensity nulls across the beam profile. A second way to control the severity of the distortion is to tune the thickness of these null intensity lines. This can easily be done by adjusting the size of the spatial filter in the 4f imaging system. We show this action in Figure \ref{fig:spatial_filter} (b), where the simulated and experimental far-field intensities of flattop beams encoded with a binary phase mask of $\Omega = 10$ are shown for different spatial filter sizes. The circular aperture size is shown by the dotted circle for clarity. This far-field profile of the beam is a physical representation of the beam's spatial frequency decomposition, with low spatial frequency components concentrated towards the centre and higher spatial frequency components located farther out. Closing the aperture decreases the size of the far-field profile, effectively cutting off a portion of the high spatial frequency components. This leaves only low frequency components to manifest in the image plane. We show these corresponding near-field profiles in Figure \ref{fig:spatial_filter} (c) for the same beams and phase mask. The simulated and experimental results both show that as the far-field profile gets smaller, the null intensity lines become larger, a result of losing more high spatial frequency components. This results in further distortion in the measured near-field intensity profile, as can be seen by the increasing $\sigma_I^2$ measured as the aperture size is decreased.

\subsection{Comparison to physical scattering samples}
\begin{figure*}[htbp]
    \centering
\includegraphics[width=0.8\linewidth]{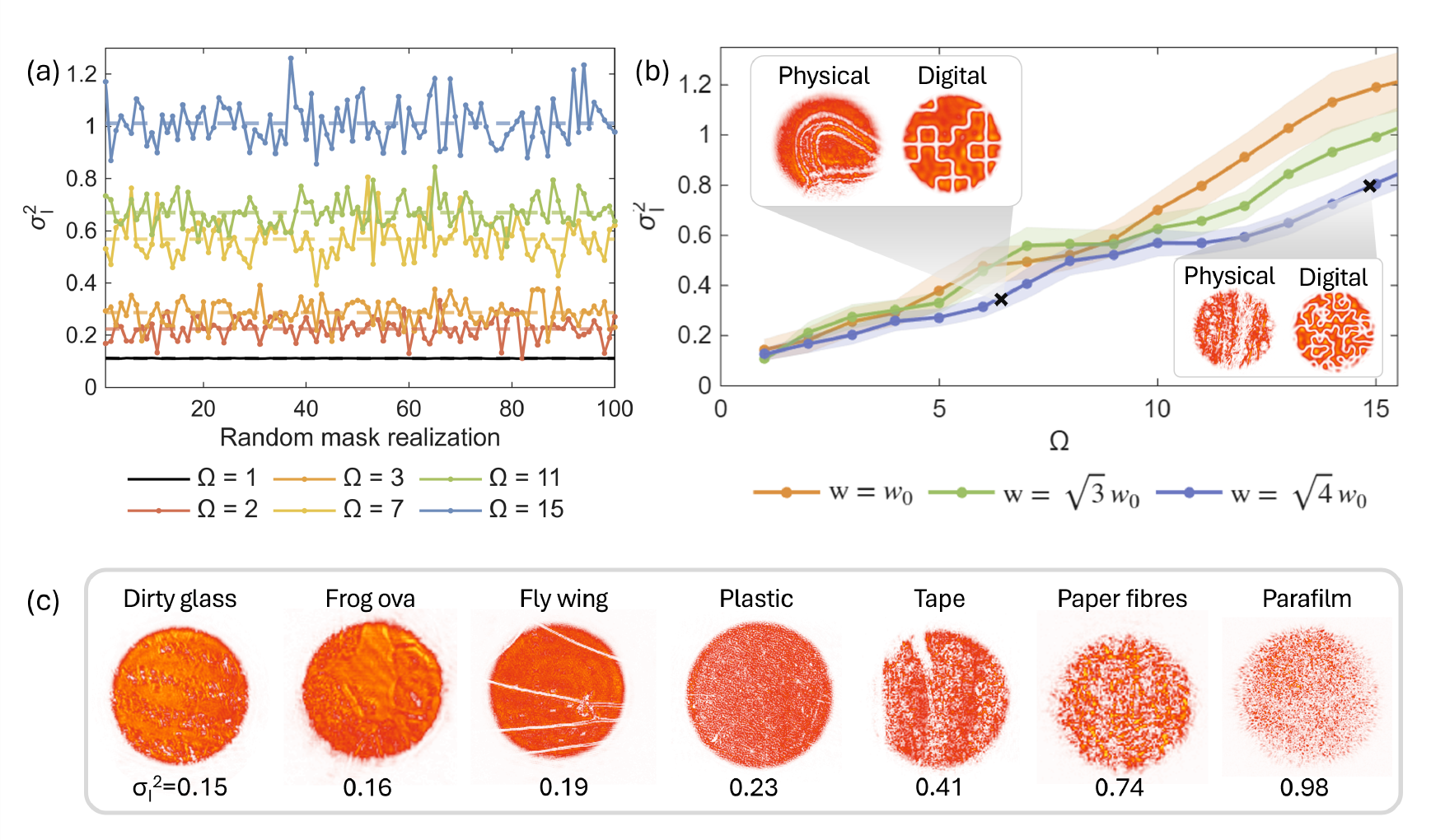}
    \caption{(a) Typical measured fluctuations in the scintillation index, $\sigma_I^2$ for 100 random realizations for different distortion strengths with mean $\sigma_I^2$ shown as dotted horizontal lines. (b) Mean and standard deviation over 100 masks for different beam radii $w$. Insets are experimentally measured digital masks, $\Omega = 6$ and $\Omega=15$ which have $\sigma_I^2 = 0.36$ and $\sigma_I^2 = 0.76$ respectively. Physical scatterers marked by black crosses show biological hydra, and bubblewrap scatterers with $\sigma_I^2 = 0.34$ and $\sigma_I^2 = 0.74$ respectively. (c) Physical samples with various $\sigma_I^2$  values accessible digitally.
    } 
    \label{fig:PW_calibration}
\end{figure*}

With precise control over the imparted distortion in hand and close qualitative agreement between theoretical/numerical and experimental results demonstrated, we turn to comparison between our digital phantom and real-world scattering samples. We again employ the scintillation index, $\sigma_I^2$, as a straightforward measure of the induced intensity fluctuations and thus the distortion severity. Figure \ref{fig:PW_calibration} (a) shows $\sigma_I^2$ for 100 random realizations of the binary phase masks at different distortions strengths $\Omega$ for a fixed aperture size of approximately $2.4$~mm. We observe that $\sigma_I^2$ is not constant for phase masks of the same $\Omega$ and seemingly fluctuates randomly about a mean value indicated by dotted horizontal lines. This means $\sigma_I^2$ increases as the encoded $\Omega $ increases, allowing us to compare the severity of our digital scatterers to real-world samples. Figure \ref{fig:PW_calibration} (b) shows the direct relationship between the mean $\sigma_I^2$ and $\Omega$ for various beam sizes $w$. The beam insets show examples of digital and real-world physical samples of similar measured $\sigma_I^2$. Several additional physical scatterers are shown in Figure \ref{fig:PW_calibration} (c) with measured $\sigma_I^2$ values all accessible with our digital phantom, showing the large range of scattering strengths both weak and strong distortion ($\sigma_I^2\sim1$).  We therefore see that our digital scattering binary phase masks perform similarly to that of real-world samples and so can be used to replicate and probe distortions of a similar nature. 

\subsection{Simulated scattering of scalar structured light}
\begin{figure*}[htbp]
    \centering
    \includegraphics[width=0.85\linewidth]{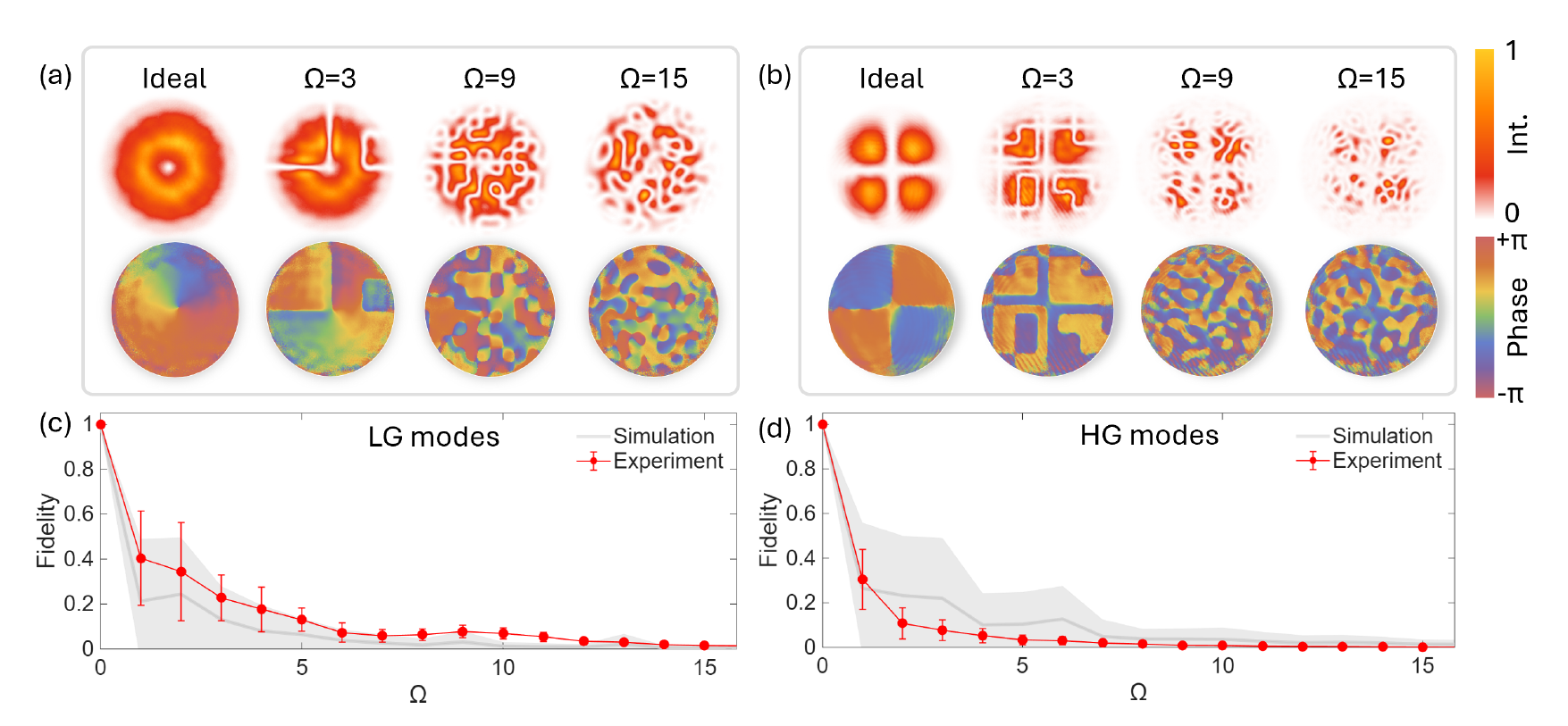}
    \caption{Measured intensity and phase profiles for various distortion strengths, $\Omega$ for (a) $\text{LG}_1^0$ modes and (b) HG modes $\text{HG}_1^1$. (c)-(d) Computed fidelity between encoded and detected modes averaged over 100 random realisations in both simulation and experiment. }
    \label{fig:LG_sim_exp}
\end{figure*}

Now that we have established that our digital scattering can indeed mimic the distortions imparted by real scatterers, we move on to apply this digital scattering to structured light modes and investigate the effects on the both the amplitude and phase. We test our digital scattering on two commonly used modal families, Laguerre-Gaussian (LG) beams imbued with orbital angular momentum and Hermite-Gaussian (HG) beams, common eigenstates of free space propagation and laser cavities. First, we investigate the digital scattering of the $\text{LG}_{\ell=1}^{p=0}$ and $\text{HG}_{m=1}^{n=1}$ modes as shown in Figure \ref{fig:LG_sim_exp}(a) and (b), respectively. As $\Omega$ increases, we see increasingly distorted near-field intensity (upper row) and phase (lower row) profiles. We quantify this increasing difference between ideal $U(x,y)$ and distorted beams $V(x,y)$ as $\Omega$ increases by computing the fidelity $F$ given by \cite{pinnell2020modal}
\begin{equation}
    F = \int_{-\infty}^{\infty} \int_{-\infty}^{\infty} U^*(x,y) V(x,y) \text{d}x\text{d}y\,,
\end{equation}
where $^*$ indicates the complex conjugate. We plot the measured fidelity as a function of $\Omega$ for the two different modes in Figure \ref{fig:LG_sim_exp} (c) and (d), respectively. We observed that the for both modes, the fidelity drops almost exponentially as $\Omega$ increases, with good agreement between the simulated and experimental results.

\begin{figure*}[htbp]
    \centering
    \includegraphics[width=0.9\linewidth]{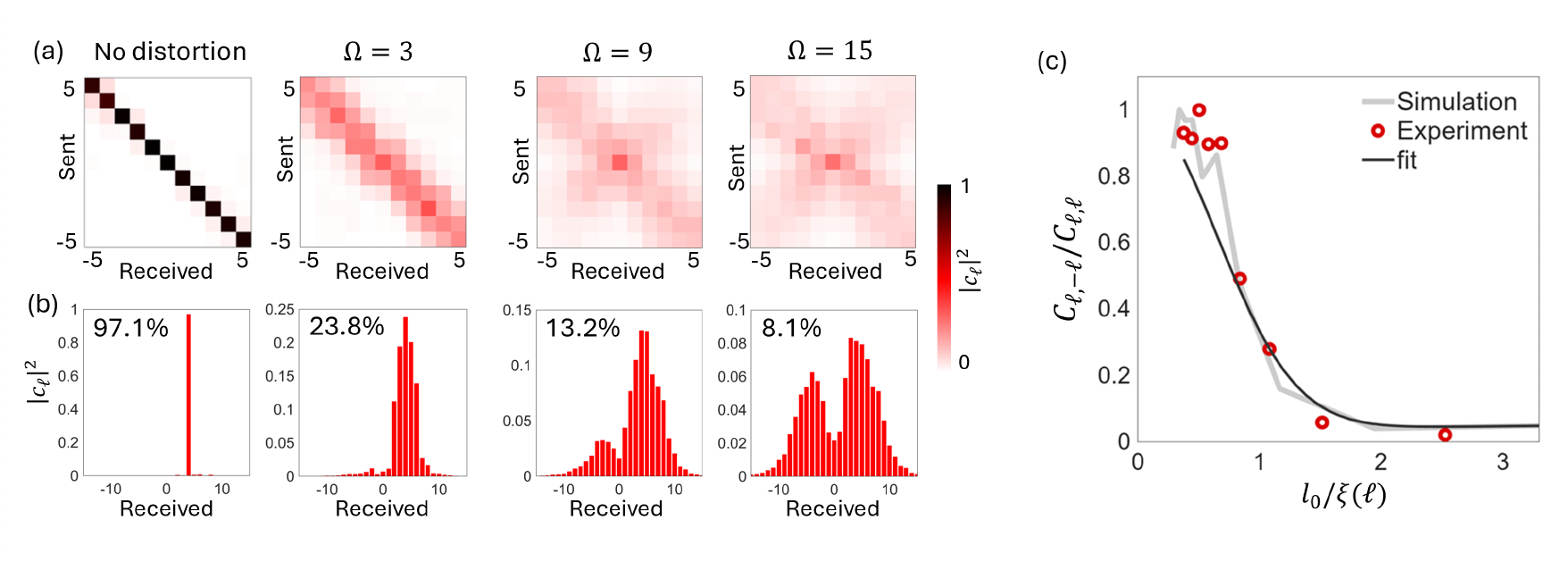}
    \caption{(a) Crosstalk matrix averaged over 100 mask realizations showing spectrum spread as $\Omega$ increases. (b) histograms for $\ell_\text{sent} = 2$ with fidelity indicated in top left. (c) Ratio of diagonal $C_{\ell,\ell}$ and anti-diagonal $C_{\ell,-\ell}$ elements of (a) for $\text{LG}_{\ell=4}$ for different distortion strengths decreasing from left to right. Fitted universal power scaling law in black.
}
    \label{fig:LGcrosstalk_powerlaw}
\end{figure*}

To further investigate this modal fidelity decay, we apply the same procedure to a range of LG modes with $\ell\in(-5, 5)$ and $p=0$ and measure the modal spread after passing through the digital phantom \cite{pinnell2020modal}. We show the resulting crosstalk matrices in Figure \ref{fig:LGcrosstalk_powerlaw} (a). We see that when no mask is applied, there is very little modal crosstalk since the transmitted beam closely matches the encoded beam. This is confirmed by the line-out taken for transmitted $\ell=2$ plotted in Figure \ref{fig:LGcrosstalk_powerlaw} (b), indicating a 97.1\% fidelity in the absence of the digital mask. However as $\Omega$ increases, the digital scatterers cause coupling between LG modes as has been observed for other complex channels \cite{cox2020structured,paterson2005atmospheric}. Figure \ref{fig:LGcrosstalk_powerlaw} (a) shows that modal crosstalk manifests not only around the diagonal but also around the anti-diagonal. The asymmetric two-peak spectrum emerging at $\pm\ell$ occurs since the LG modes of opposite sign are identical in intensity, with similar effects seen in simulations of Kolmogorov turbulence and Gaussian noise \cite{Bachmann2024universal}. 

Figure \ref{fig:LGcrosstalk_powerlaw} (c) shows the experimental and simulated ratio of the diagonal, $C_{\ell,\ell}= |c_{\ell,\ell}|^2$, and anti-diagonal elements $C_{\ell,-\ell}$  versus encoded correlation length $l_0$ rescaled by the phase correlation length of the $LG_\ell^0$ beam, $\xi(\ell)$ calculated as described in Ref. \cite{leonhard2015universal}. We fit to this the universal, $\ell$-independent crosstalk law for twisted light in complex media \cite{Bachmann2024universal} shown by the black line given by $f(x) =(1-a) e^{-x^2/(2b^2)}+a -cx$, with free fitting parameters $a = 0.035$, $b = 0.65$ and $c = -0.004$. The close agreement between simulation, experiment and analytical fit function for the shown example of $\ell=4$ further confirms the validity of our digital scattering approach for simulating real-world scattering channels and using this approach for structured light applications. 

\subsection{Simulated scattering of vectorial structured light}
\begin{figure*}[htbp]
    \centering
    \includegraphics[width=0.8\linewidth]{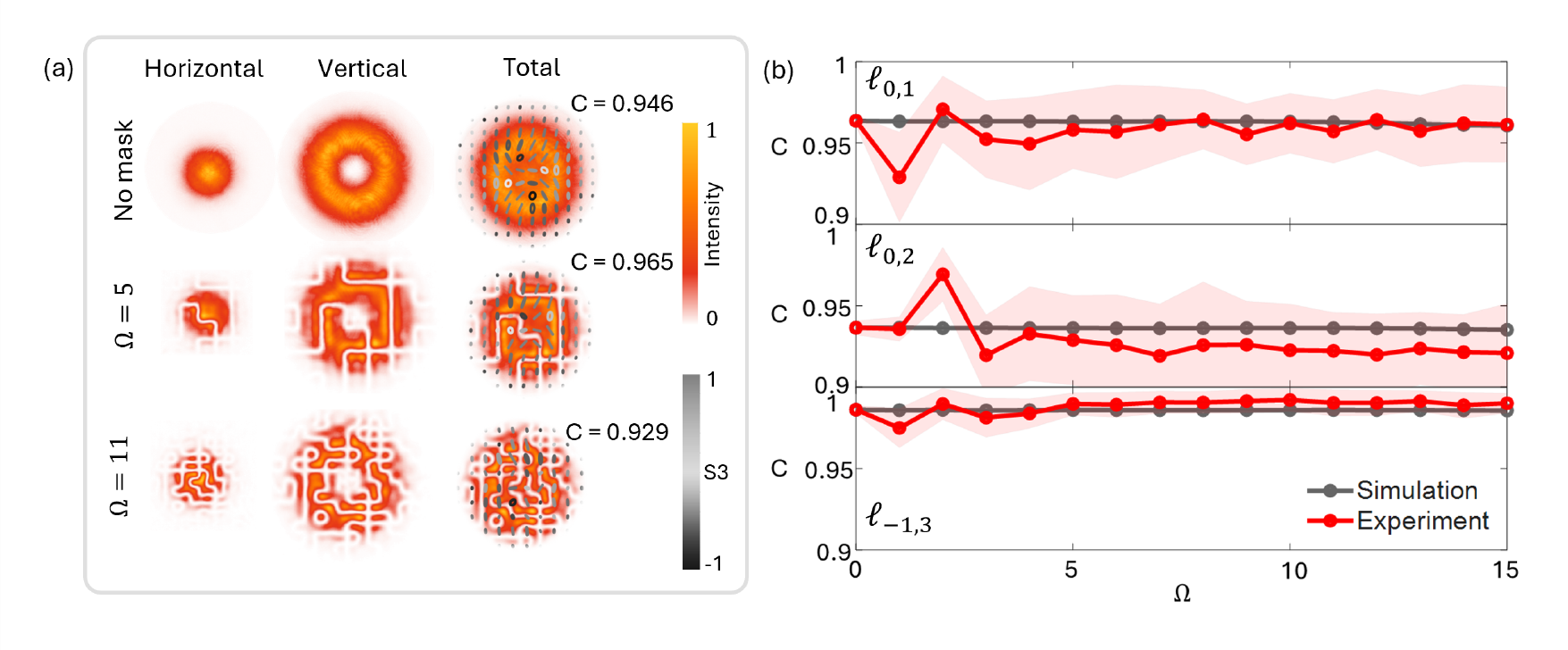}
    \caption{(a) Measured horizontal and vertical component LG modes with $\ell_1=0$ and $\ell_2=2$ and the total intensity profile of the vector beam with polarization ellipses. Inset indicates the measured concurrence, $C$. (b) Concurrence versus distortion strength, $\Omega$ for different vectorial beams generated from LG beams with different azimuthal indices $\ell_1, \ell_2$ as shown, averaged over 100 random mask realisations at each distortion strength. 
    \label{fig:vectorbeams}
    }
\end{figure*}

As a final application, we investigate the effects of our digital phantom on vectorial structured light with spatially varying polarisation generated using superpositions of orthogonally polarized LG modes. Figure \ref{fig:vectorbeams} (a) shows the measured intensity projections of horizontal and vertical polarizations, as well as the total intensity and state of polarization for a beam with $\ell_1 = 0$ and $\ell_2 =2$. The same phase mask is applied to both polarizations on the SLM to emulate a non-birefringent physical scatterer. As expected, we see similar distortion to the both the vertically and horizontally polarised components. Importantly for  vectorial light, we also observe that the spatially varying polarisation structure is also distorted, with it becoming more scrambled as $\Omega$ increases.

The masks we used are phase-only, and as such represent a unitary channel that should not alter the beam's degree of non-separability \cite{nape2022revealing}. To verify this, we compute the concurrence $C$ of the beam \cite{selyem2019basis} (a quantitative measure of its degree of non-separability) with an initial value of $C=1$ for various $\Omega$ and plot the results in Figure \ref{fig:vectorbeams} (b). We observe only a minimal drop in concurrence, likely due to lost spatial frequencies in the imaging system as discussed previously. However, we observe that the concurrence is extremely stable across all distortions strengths, with little deviation in the mean value of $C$ for all $\Omega$ and for all three beam types tested.

%s $\Omega$ increases, the increasing intensity nulls decrease the total power in each polarization. However since the relative total power between horizontal and vertical polarizations with different $\Omega$, remains the same with only small variations due to the random mask realisation. This occurs for various vectorial beams shown in Figure \ref{fig:vectorbeams}(b), with fluctuating concurrence values around a mean value due to averaging over 100 mask realisations.

%, denoted by as $P_H$ and $P_V$ and shown in Figure \ref{fig:vectorbeams}(b). For clarity of comparison their powers are normalised such that  $P_H =P_V=1$ at $\Omega=0$. The intensity nulls cause a decrease in power as the distortion strength increases reaching as low as only 50\% of the original power remaining at $\Omega=15$.

%This is also consistent with the non-unitarity of such a lossy channel \cite{nape2021revealing} which emulates real-world scatterers which typically direct light away from the detector []. Figure \ref{fig:vectorbeams}(c) shows the changing concurrence values more explicitly as a function of $\Omega$ for various vector beams. Notably, as distortion strengths increase toward very severe $\Omega >17$, the power in both beams becomes very low such that $P_H>P_V$ and so the measured concurrence increases at large $\Omega$.

\section{Extensions and alternative implementations}
\begin{figure*}[h]
    \centering
    \includegraphics[width=0.8\linewidth]{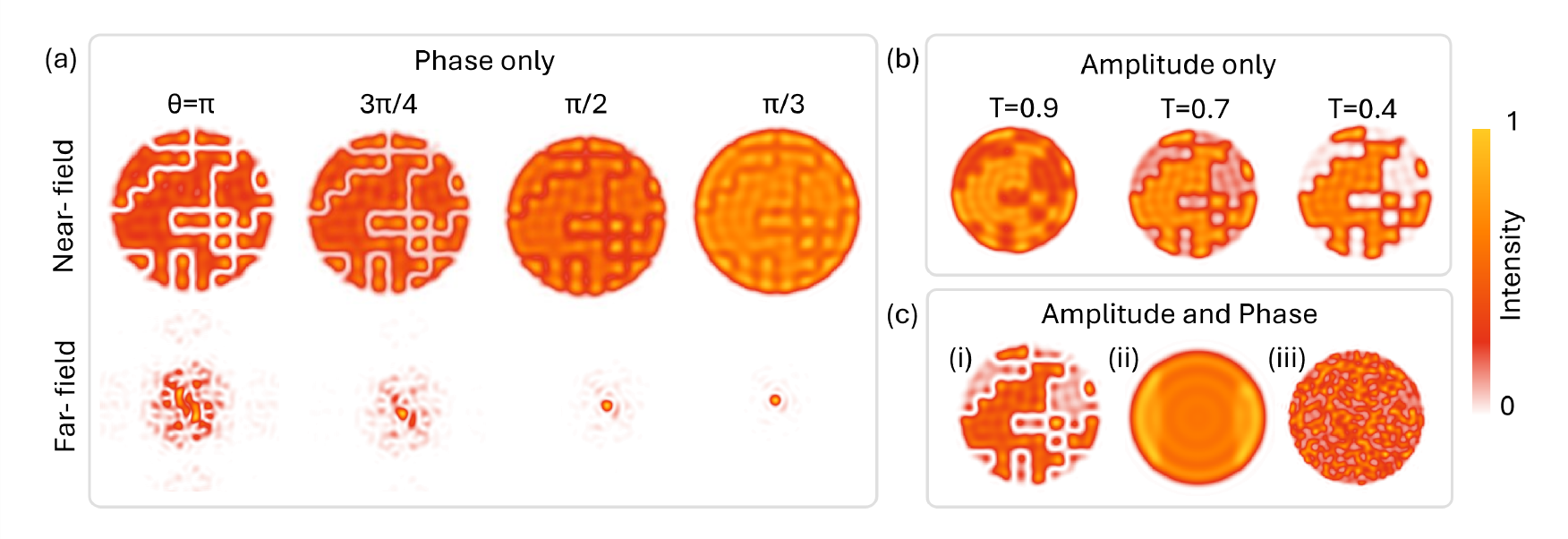}
    \caption{(a) Effects of phase-only masks in the the near and far-field intensity patterns when $\Omega=8$, for various maximum phase values, $\theta$. (b) Near field effect of amplitude only masks, where only a fraction of the intensity is transmitted, $T$ in certain superpixels for $\Omega=8$. (c) Near field effect of amplitude and phase distorting masks with varied parameters: 
    (i) $\Omega=8 \text{ },\theta=\pi,\text{ } T=0.7$, 
    (ii) $\Omega=2 \text{ },\theta=\pi/6,\text{ } T=0.98$, 
    (iii) $\Omega=35 \text{ },\theta=\pi/5,\text{ } T=0.75$. All images are numerically simulated.
    } 
    \label{fig:alternatives}
\end{figure*}

For simplicity, the digital scatters demonstrated thus far used binary $0$ and $\pi$ phase masks with square super-pixel geometry. However, this approach is easily extended to more general conditions depending on the user's application of interest.  To illustrate this, consider the examples in Figure \ref{fig:alternatives}. Figure \ref{fig:alternatives} (a) shows the effect of varying the maximum phase imparted, $\theta$, both in the near and far-field. These masks with smaller phase shifts have fewer high frequency components as shown in the far-field profiles. Consequently $\theta$ can be used to tune the total loss in the intensity nulls. Alternatively a user could easily combine these $\theta$ masks intelligently to implement random spatially varying and multi-level phase only masks to mimic materials of more smoothly spatially varying refractive index or thickness. The second, and perhaps most natural extension to our digital phantom, is simply to perform all experiments in the far field rather than the near-field. Figure \ref{fig:alternatives} (a) highlights that the far-field profiles are also random and are tuned by both $\Omega$  and $\theta$ since phase is naturally imparted into randomness in the intensity profile. Third, absorption and amplitude distortion could also be induced by these masks by encoding some loss on the SLM hologram, easily achievable through a complex amplitude encoding scheme already implemented to generate the various formas of structured beams in this work \cite{arrizon2007pixelated}. A few examples are shown in Figure \ref{fig:alternatives} (b) where the total fraction of transmitted intensity, $T$ in certain superpixels is easily controlled. Finally, these absorption and phase effects can be combined allowing for a wide variety of digital phantoms with some examples shown in Figure \ref{fig:alternatives} (c). These range from sharply distorted features in (i), to smoothly varying weak distortions in (ii) or rapid phase fluctuations without much change in amplitude in (iii) by combining various $\Omega,\theta$ and $T$ parameters. While not exhaustive, these examples serve to convey the power and versatility of the approach.  Additionally, the use of square blocks for the phase mask need also not be the case.  We elected square blocks for ease of use on SLMs, but the super-pixels can encode arbitrary shapes. Furthermore, the masks encoded on the hologram could be rotated or shifted randomly to avoid anisotropies \cite{Bachmann2024universal}. Including one or more of these effects to the basic cases we have shown here could considerably increase the capabilities of such a digital phantom depending on the needs of the user. 

 Polarisation-dependent effects and birefringence may also be emulated using our approach by making careful use of our vectorial light generation setup or similar. A Sagnac interferometer allows one to tune the amplitude and phase profile of each polarisation component, completely independently. This allows one to impart different phase delays or different amplitude distortions into each component. Finally, in all tests we ran 100 random samples of a given scattering strength as sequential holograms.  Although our interest was in time averaged results, the approach is easily extended to studying single-shot time varying media by synchronising the medium with the hologram display rate.  The SLM allows this to be done at 100s of Hz, while digital micro-mirror technology can push this to tens of kHz rates. 

\section{Conclusion}
We have presented and experimentally demonstrated a simple technique to generate highly controllable digitally simulated randomness and scattering. Our approach relies only on a single binary phase screen encoded onto an SLM, making it a functional and readily accessible experimental tool for experiments only requiring an SLM and basic optical components. To demonstrate the versatility of our approach, we showed two methods to tune the distortion strength and characterised their effects. Next we applied our digital scattering phantom to three common forms of structured light, uniform plane waves, scalar LG and HG modes and vectorial beams. In each case we showed the effects on the output amplitude and/or phase and benchmarked the results in reference to physical inorganic and biological scatterers, showing results consistent with previous findings for structured light in other complex channels. In all cases we observed excellent agreement between simulation and experiment. Finally, we provide several alternative approaches and extensions to this basic digital phantom allowing the user to tailor the digital phantom to their requirements and opening the way to more sophisticated digital phantoms facilitating more systematic and controlled studies of structured light in random and complex media. 

\section*{Disclosures}
The authors declare no conflicts of interest.

\section*{Data Availability}
Data underlying the results presented in this paper are not publicly available at this time but may be obtained from the authors upon reasonable request.

\begin{acknowledgments}
C. Peters and A. Forbes acknowledge funding from the National Research Fund/CSIR Rental Pool Program and the Oppenheimer Memorial Trust.
\end{acknowledgments}

\section*{Data Availability Statement}

The data that support the findings of this study are available from the corresponding author upon reasonable request.

\end{document}